# Dynamic Radio-Frequency Transverse Susceptibility in Magnetic Nanoparticle Systems

L. Spinu*, H. Srikanth, E. E. Carpenter and C. J. O'Connor
*Advanced Materials Research Institute, University of New Orleans, New Orleans, LA 70148*

A novel resonant method based on a tunnel-diode oscillator (TDO) is used to study the dynamic transverse susceptibility in a Fe nanoparticle system. The magnetic system consists of an aggregate of nanometer-size core (Au)-shell (Fe) structure, synthesized by reverse micelle methods. Static and dynamic magnetization measurements carried out in order to characterize the system reveal a superparamagnetic behavior at high temperature. The field-dependent transverse susceptibility at radio-frequencies (RF), for different temperatures reveals distinct peak structure at characteristics fields (± $H_K$, $H_C$) which changes with temperature. It is proposed that relaxation processes could explain the influence of the temperature on the field dependence of the transverse susceptibility.) on the MI.

## I. INTRODUCTION

Since 1909, when for the first time Gans [1] discussed the notion of transverse susceptibility ( $c_T$ ), this has remained an interesting topic in magnetism. The primary importance of $c_T$ arises from its possibility to measure the effective anisotropy field in fine particle systems and to investigate the distribution in particle orientation and anisotropy fields. Both, experimental and theoretical studies were carried out in order to investigate the transverse susceptibility. After the theoretical treatment of Aharoni et al. [2] based on Stoner-Wohlfarth model [3], Pareti et al. [4] provided the first experimental confirmation through low frequency AC susceptibility measurements. The predicted peaks at ± $H_K$ and the coercive field ($H_C$) in the field variation of the transverse susceptibility were observed for a ferromagnet having a positive uniaxial anisotropy. More recently, Chantrell et al. [5], and Hoare et al. [6] have pointed out the importance of orientational texture on the transverse susceptibility of particulate recording media. It should also be noted that transverse susceptibility plays an important role in related new phenomena with potential applications, like Giant Magneto-Impedance (GMI) [7].

In this paper we propose a new experimental method to probe the transverse susceptibility using a tunnel-diode oscillator (TDO) technique operating at a frequency around 5 MHz. In all previous treatments of the transverse susceptibility in particulate magnetic systems, relaxation phenomena were neglected. Here we present our results on a nanophase magnetic system where magnetic relaxation must be taken into account. To the best of our knowledge, these are the first experimental results of the field variation of $c_T$ for various temperatures. In our experiments, with the characteristic experimental time related to the frequency of the RF magnetic field ($t_m = 1/n$), by varying the temperature we can outline very clearly the blocked and superparamagnetic states. These results could be interpreted by taking into account the relaxation effects.

## II. EXPERIMENTAL

The system we studied consists of nearly uniform spherical nanoparticles having an Au-Fe-Au structure with a gold core, ring shaped annular iron nanostructure surrounded by a gold shell. This nanophase magnetic system was synthesized by a sequential reverse micelle method described elsewhere [8]. The fine particle comprises of a gold core, which acts as a nucleation source for the epitaxial growth of an iron shell. The iron shell is passivated using a second gold shell. Structural characterization of the system done using X-ray diffraction and electron microscopy [10] confirms the core-shell structure with an average diameter for the magnetic particles of 8nm.

Magnetic characterization was performed with a commercial superconducting quantum interference device (SQUID) magnetometer. The highest applied field was 5.5T and the lowest temperature was 2K. Figure 1 shows the temperature dependence of the magnetization and the inset shows the hysteresis loop at 10K displaying a coercivity of 400 Oe.

The magnetic behavior displayed is typical for a fine particle system where the relaxation of the magnetic moment of the particles is present. In the ZFC process, at the lowest temperatures, the total magnetic moment is nearly zero due to freezing of the moments in random directions. As the temperature is raised, the total magnetic moment increases due to the gradual alignment of moments in the field direction until it reaches the maximum value (around 70K). The maximum in the ZFC curve arises due to competition between an increase in the number of particles oriented in the field direction and decrease in the susceptibility of each particle. The broadness of the maximum of the ZFC curve can be related to the size distribution of the nanoparticles but is also a sign of the presence of interactions between them. It should be noted that in powders, the dipolar interaction between magnetic moments could be significant.

* Corresponding author; E-mail: LSPINU@UNO.EDU

To probe the dynamic transverse susceptibility in the radio-frequency range we employed a very sensitive method that we have recently developed [9]. Our experiments are based on a novel tunnel-diode oscillator (TDO) technique that combines the advantages of the precision afforded by resonance frequency measurements along with the transverse modulation principle. The measurements were done using a LC- tank resonator driven by a tunnel diode forward biased in its negative resistance region. This idea is the basis of the TDO measurement system whose circuit design and operation have been described elsewhere in detail [9]. The circuit is self-resonant with a typical resonance frequency around 5MHz.

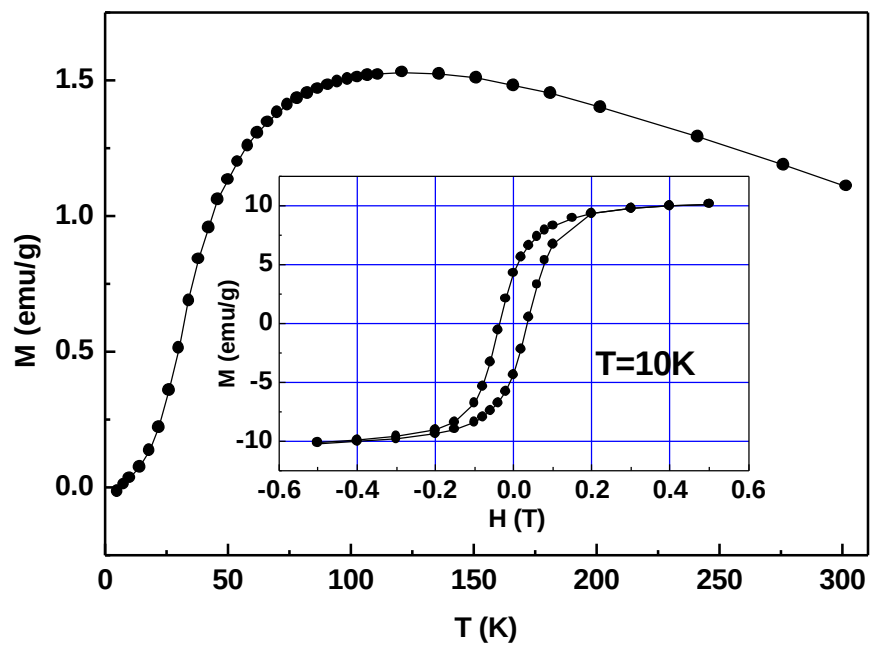

Fig.1 Temperature dependence of the magnetization of the sample for a ZFC process in a magnetic field of 50 Oe. Inset- hysteresis loop at 10K.

The sample in powder form is placed in a gelcap that snugly fits into the core of an inductive copper coil *(*L). This is inserted into the sample space of a commercial Physical Property Measurement System (PPMS) from Quantum Design using a customized radio-frequency (RF) co-axial probe. The temperature and static magnetic field are varied using the PPMS. The oscillating RF field $H_{rf}$ produced by the RF current flowing in the coil windings, is oriented perpendicular to the static field and this arrangement sets up the transverse geometry. In the experiment, the measured quantity is the shift in resonant frequency as the static field is varied. The frequency shift arises from a change in coil inductance L, determined by the change in transverse permeability $m_t$ of the sample. Thus, knowing the precise geometrical parameters we can derive an absolute value for the transverse susceptibility $c_t = m_t - 1$. The transverse susceptibility ratio can be written as:

$$\Delta c_t / c_t(\%) = \frac{[c(H) - c_{sat}] \times 100}{c_{sat}} \qquad (1)$$

where $c_{sat}$ is the transverse susceptibility at the saturating field $H_{sat}$=1.5T. This quantity which represents a figure of merit, does not depend on geometrical parameters and is useful for comparing the transverse susceptibility data for different samples, or for the same sample under different conditions like varying temperature.

## III. RESULTS AND DISCUSSION

The transverse susceptibility ratio ($\Delta c_t / c_t$) for different temperatures, measured in field sweeps from positive (+1.5T) to negative saturation (-1.5T) is shown in Fig. 2. A prominent feature that can be distinguished clearly (as seen in inset of Figure 2), is that for low temperatures (below ~50K) the curves show two peaks located symmetric about the origin of the field axis. Also, these peaks have different heights, with the larger peak located at a positive field when the field is ramped from positive to negative values. As the temperature is increased, this double-peak structure becomes less pronounced and merges into a single central peak. Also with increasing temperature, the heights of the two peaks become equal for *T*~50K. All the $\Delta c_t / c_t$ curves are more flattened exhibiting a larger width for lower temperatures than for higher ones.

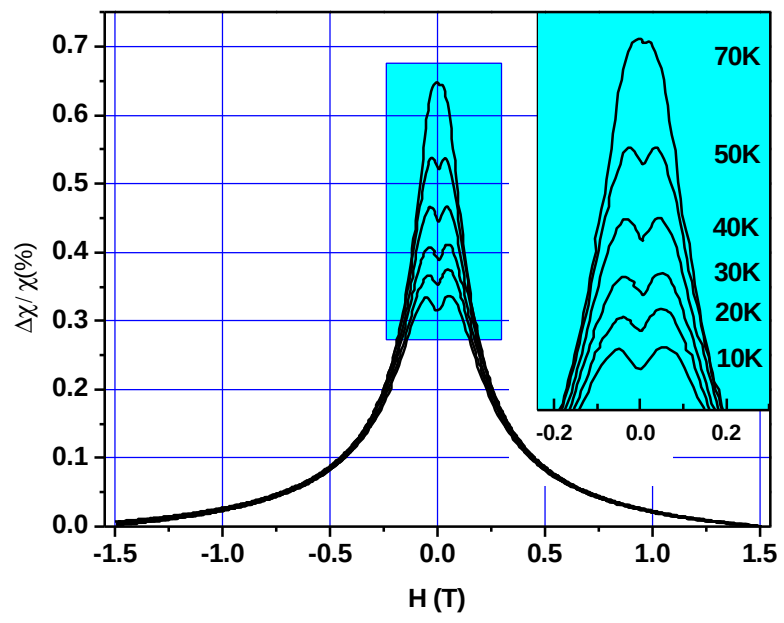

Fig. 2 Transverse susceptibility ratio for various temperatures.

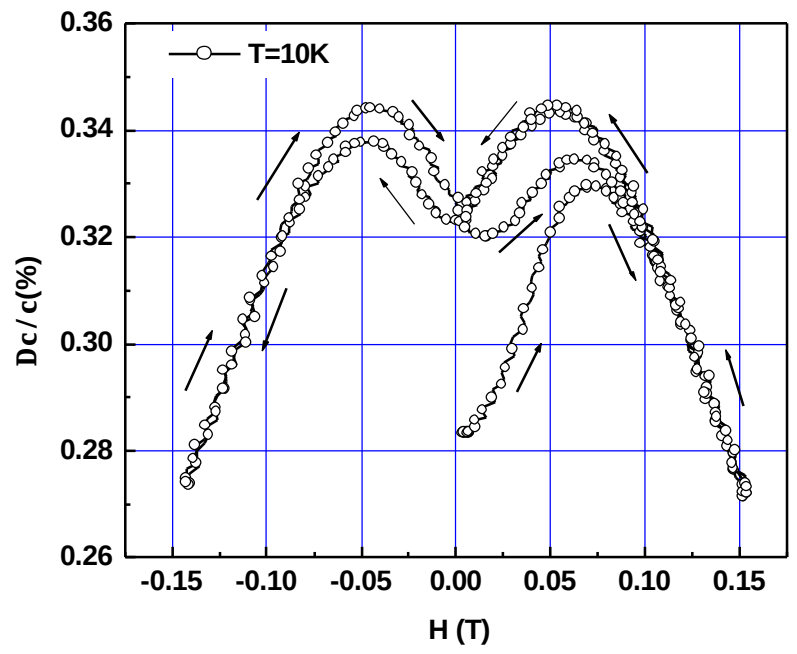

Fig.3. Low-field hysteresis and peak structure in the transverse susceptibility for T=10K

Cycling the magnetic field with the sample held at low temperatures, we observe an irreversible behavior. To map out the low-field hysteresis and study the peaks in finer detail, we started at zero field and cycled the magnetic field up and down between –1.5T and +1.5T. The transverse susceptibility ratio for $T$=10K is plotted in Fig.3. The peak structure and hysteresis systematics are now clearly visible. As the field is varied over the entire loop, two observations can be made about the peak structure: a) symmetric location around ±0.05T and b) asymmetric peak heights as the field is swept in one direction from –0.15T to +0.15T or vice versa (follow arrows in Fig. 3). From the known models of transverse susceptibility, these peaks at ±0.05T are related to the effective anisotropy field $\pm H_k$ and their rounded nature is consistent with the presence of a distribution of the anisotropy fields.

These features in the field variation of the transverse susceptibility in the low temperature domain are somewhat similar to those observed by Pareti and Turilli [4] on micron-sized particulate $BaFe_{12}O_{19}$ and by A. Hoare et al. [6] on other fine particle oxide systems. However in our case, we do not observe clearly the presence of the three peaks (viz. two symmetric ones associated with the anisotropy field and one with the coercive field) for a single field sweep from positive to negative saturation (or vice versa). This is due to the fact that in our case the anisotropy and coercive fields are very close and the two maxima are superposed. Moreover, the rounded nature of the peaks is consistent with the presence of an anisotropy field distribution that makes the coercive peak indistinguishable. The theory due to Aharoni [2] based on the coherent rotation model [3] predicts the existence of the peaks at these characteristic fields.

The experimental results observed in our nanoscale system for higher temperatures ($T$ > 70K) does not contain the irreversible behavior observed at low temperatures. This can be understood only if we take into account the effect of the relaxation of the particle magnetic moments. The reasoning can be made as follows. At low temperatures, particle moments reach their equilibrium positions determined only by the value of the applied DC magnetic field and its orientation with respect to the easy axis. So, in this case, the switching from one orientation to the other is determined only by the magnetic field. As the temperature is increased, the thermal energy can determine the transition from one equilibrium position to the other, even at low magnetic fields. This eliminates the need for a switching field and leads to a reversible behavior and also accounts for the vanishing of the two-peak structure.

It should also be pointed out that the transverse susceptibility models are based on standard Stoner-Wohlfarth theory which applies only for a non-interacting system. In real cases, nanoscale materials in powder form invariably have inter-particle interactions. So, to completely model our data, modifications to the Aharoni theory is needed that takes into account the effect of relaxation and interactions. This is a complex problem that may be numerically addressed by including a distribution in interaction fields (similar to the Preisach model [11,12]) in the transverse susceptibility calculations. We are currently developing such a model and a complete analysis will be presented in a separate publication.

## IV CONCLUSION

In conclusion, we have demonstrated an effective method to study the dynamic transverse susceptibility in magnetic nanoparticle systems using a sensitive resonant experimental technique. Characteristic peaks at anisotropy fields are observed and their temperature evolution shows a systematic trend. We identify the important roles of relaxation and interaction effects in dynamic transverse susceptibility experiments on nanoparticle systems.

## ACKNOWLEDGEMENTS

This research is supported by DARPA through grant No. MDA 972-97-1-0003. The authors would like to thank Jason Wiggins for assistance with RF instrumentation.